\documentstyle[12pt]{article}

\small
\begin{document}
\pagestyle{myheadings}

\title{
{\bf Two-phonon states in alkali-metal clusters. }
}
\author{ {\bf Abdellatif Abada}\thanks {email : "ABADA@ipncls.in2p3.fr"}
{\bf ~and Dominique Vautherin}\thanks {email : "VAUTHERIN@ipncls.in2p3.fr"}
\\
{\normalsize
Division de Physique Th\'eorique,
\thanks{Unit\'e de Recherche des Universit\'es Paris XI
et Paris VI associ\'ee au C.N.R.S } }\\
{\normalsize	Institut de Physique Nucl\'eaire,} \\
{\normalsize	F-91406 , Orsay Cedex, France }
}
\date{}
\maketitle
\begin{abstract}
Two phonon-states of alkali-metal clusters (treated as jellium spheres) are
calculated by using a method based on a perturbative construction of periodic
orbits of the time-dependent mean-field equations. Collective vibrations with
various multipolarities in charged $Na^+_{21}$ are considered.

\end{abstract}


\vskip 3.5cm
{\it Submitted to Phys. Rev. B }
\vskip 1cm
IPNO/TH 92-80 \hfill{ December 1992}

\newpage
\begin{center}  {{\large {\bf INTRODUCTION }} } \end{center}

For metallic clusters, it appears reasonable to approximate the valence
electrons as a system of independent fermions enclosed in a box,
in close analogy with the mean-field description of nucleonic motion in nuclei.
The many body methods developped for collective modes in nuclei
are thus a natural tool to investigate, in a first stage, collective motion in
metallic clusters.
The length and the energy scales in metallic clusters and nuclei are of course
widely different. The interactions also have a different nature.
The strong force mediated by pion exchanges ensures nuclear binding while
electrons interact with each other and with the nuclei through electromagnetic
forces.

The analogies between nuclei and metallic clusters became apparent in 1984
with the experimental discovery
of shell structure and magic numbers in sodium clusters by W. D. Knight {\it et
al} \cite{1} and parallel theoretical efforts by W. Ekardt \cite{2}.
Since then, the physics of metallic clusters has been of continuing interest
\cite{3}. The valence electrons in these clusters can be
described as delocalized particles moving in a mean-field due to their mutual
interaction and to a uniform positive background.
This approximation is known as the jellium model. As a result, the
ionic background contributes only an electrostatic attraction to
the mean-field, whereas the valence electrons are treated
as quantum particles exhibiting exchange and correlation effects in addition to
their electrostatic repulsion. This treatment is usually carried out in the
framework of the density-functional formalism \cite{4,5}.

The experimental discovery of a giant dipole resonance in metallic clusters
interpreted as a collective electronic excitation (surface plasmon)
\cite{6,7,8} induced a substantial theoretical effort.
To understand the presence of collective motion, the usual
random-phase approximation approach (RPA) and the time-dependent
local-density approximation (TDLDA) have been applied [9-14].
Our aim in this article is to describe collective motion of large amplitude
in metallic clusters with a method which is more general than RPA.
It is based on secular perturbation theory. More precisely we start from
a perturbative method to construct finite-amplitude periodic orbits of the
nonlinear time-dependent mean-field equations.
This method was developped recently to describe giant collective resonances
in nuclear physics \cite{15,16} and the Skyrmion breathing mode in
hadronic physics \cite{17}. Although this method is not
always applicable due to the possible presence of branches and bifurcations
\cite{23} we found it well adapted to investigate giant collective
oscillations in the case of nuclei. From the knowledge of periodic orbits we
were able to calculate collective
energy spectra by a semiclassical quantization or by a construction of the
collective Bohr-type Hamiltonian.

The present article is organized as follows. Sec. I presents a summary of
the self-consistent spherical jellium-background model and defines our
notation.
In Sec. II we review briefly the expansion method of Refs. \cite{15,16,17}
and we specialize to the alkali-metal clusters case.
In Sec. III we present our results concerning monopole, dipole, quadrupole and
octupole modes in charged $Na^+_{21}$ (closed-shell spherical cluster with
20 electrons), and perform a comparison with available data \cite{177,178} and
other theoretical results \cite{9,11,18}.
In Sec. IV we discuss the applicability of the method to metallic clusters and
summarize our main conclusions.

\renewcommand{\theequation}{\arabic{section}.\arabic{equation}}
\setcounter{equation}{0}
\section {The jellium model}

 In the jellium model, the $N$ ions are replaced by a uniform background of
positive charge having radius $R = r_s N^{\frac{1}{3}} $ where $r_s = 3.93$
a.u.
for sodium. In the density functional approach the electrons gas energy
$E[\rho]$ is assumed to be a function of the one-body electron density matrix
$\rho(t)$
\begin{equation} \label{1.1}
\langle {\bf r}\vert \rho(t) \vert {\bf r'} \rangle = \sum_{i=1}^A
\phi_i({\bf r},t) \phi^*_i({\bf r'},t) ,\end{equation}
where the $\phi_i$'s are the single particle wave functions of the $A$
occupied single electrons
orbitals. In this work, we use the density functional (see Ref. \cite{4} for a
general discussion )
\begin{equation} \label{1.2} \begin{array}{ll}
E[\rho] ~=~ &{\displaystyle \int \hbox {d}^3 r ~\Big \{ ~
\frac {1}{2} \tau ({\bf r},t) ~-~
\frac {3}{4} (\frac{3}{\pi})^{1/3} \rho^{4/3} ({\bf r},t) ~+~
\rho({\bf r},t) {\cal E}_{xc} (\rho)  ~+
{}~\frac {1}{2} \int \hbox {d}^3 r' ~\frac{\rho({\bf r},t) \rho ({\bf r'},t)}
{\vert {\bf r} - {\bf r'} \vert}  ~+} \\
&{\displaystyle
 V_J({\bf r}) \rho({\bf r},t) ~\Big \}
{}~+ E_{JJ}  }  ~~. \end{array} \end{equation}
We use atomic units $(e=m_e=\hbar=c=1)$ for which energies are expressed
in units of
$2R_y = 27.2$ eV ($R_y$ being the Rydberg constant) and lengths in units of
the Bohr radius $a_0 = 0.53 \AA$. In Eq. (\ref{1.2}) the first term is the
kinetic energy density
$\tau({\bf r},t) = \sum_{i=1}^A \vert \nabla \phi_i({\bf r},t) \vert^2 $,
the second term is the Coulomb exchange in the local density approximation.
The third one simulates correlations, while the fourth one is the direct
Coulomb term. The fifth term is the Coulomb interaction of the electrons with
the jellium. For an uniform density
distribution for the $N$ ions, the jellium coulomb energy $E_{JJ}$
is $\frac {3}{5} N^2/R$ and $V_J$ is given by
\begin{equation} \label{1.3}
V_J (r) =  \Bigg \{ \begin{array} {ll}
-N/r ~~~r>R \\ {\displaystyle
-\frac {3}{2} \frac {N}{R^3} (R^2 - \frac {r^2}{3} ) ~~~r<R ~}.
\end{array} \end{equation}
For the correlation term, we follow the Gunnarsson-Lundqvist parametrization
\cite{5} :
\begin{equation} \label{1.4}
{\cal E}_{xc} = {\displaystyle
- 0.0333 ~G[\frac {1}{11.4} (\frac{3}{4\pi \rho} )^{1/3} ] } \end{equation}
where the function $G$ is defined by
$$G(x) = (1+x^3) \hbox {Ln} [1+1/x] - x^2 + x/2 -1/3 ~.$$
The time-dependent mean-field equation reads \cite{20}
\begin{equation} \label{1.5}
\hbox{i} \dot \rho = [W(t),\rho(t)] ~,\end{equation}
where $W(t)$ is the mean-field Hamiltonian $\delta E/\delta \rho $. It reads
$-\frac{1}{2} \Delta + U({\bf r},t)$ with
\begin{equation} \label{1.6}
U({\bf r},t) = {\displaystyle
- (\frac{3}{\pi})^{1/3} \rho^{1/3} ({\bf r},t) - 0.0333 ~\hbox {Ln}
[ 1 + 11.4 (\frac {4\pi}{3} \rho({\bf r},t) )^{1/3} ]
+ \int \hbox {d}^3 r' ~\frac{ \rho ({\bf r'},t)}
{\vert {\bf r} - {\bf r'} \vert}  + V_J(r) } \end{equation}
Finally, we recall that the one-body density matrix defined in Eq. (\ref{1.1})
must satisfy at all times the conditions $\hbox {Tr} \rho(t) = A$
($A$ being the number of electrons) and $ \rho^2(t) = \rho (t)$ \cite{20}.

At this stage a word of caution about the application of the variational
procedure should be given. The energy functional should
be minimized with respect to the electron density only. In contrast it
should not be minimized with respect to the jellium parameters.
Indeed, such a procedure would lead to the prediction that potassium and
sodium clusters have the same radius, which is in conflict with observations.
This difference also suggests that it is not legitimate to perform a
minimization with respect to the shape of the jellium.

\renewcommand{\theequation}{\arabic{section}.\arabic{equation}}
\setcounter{equation}{0}
\section { The method }
The method we use to construct periodic orbits of the mean-field
equation (\ref{1.5}) with a given period $T$ consists in expanding the one-body
density matrix $\rho(t)$ as a power series in the amplitude of the vibration
$\epsilon$ \cite{16}. In order
to take into account the dependence of the frequency $\omega = 2\pi/T$
on the amplitude \cite{19}, we expand $\rho(t)$ as
\begin{equation} \label{2.1}
\rho(t) = \rho_0 + \epsilon \rho_1(\frac {\omega}{\omega_0}t)
+ \epsilon^2 \rho_2(\frac {\omega}{\omega_0}t)
+ \epsilon^3 \rho_3(\frac {\omega}{\omega_0}t) + \dots
\end{equation}
with
\begin{equation} \label{2.2}
\omega = \omega_0 + \epsilon \omega_1 + \epsilon^2 \omega_2 + \dots
\end{equation}
The first order equation is solved by the usual ansatz \cite{20}
\begin{equation} \label{2.3}
\rho_1(t) = \eta \exp{(-i\omega_0 t)} + \eta^+ \exp{(+i\omega_0 t)} ~.
\end{equation}
After inserting the expression of $\rho(t)$ given above in Eq. (\ref{1.5}), we
obtain the evolution equations of the successive terms $\rho_k ~(k=0,1,2,..)$
by identifying the various powers of $\epsilon$ :
\begin{equation} \label{2.4}
\hbox{i} \dot \rho_k = \sum_{j+m=k} \Big \{ [W_j,\rho_m] -
\hbox{i} \Omega_j \dot \rho_m  \Big \} ~.\end{equation}
In Eq. (\ref{2.4}), the operator $W_j$ corresponds to the $j$-th order
contribution to the
mean-field Hamiltonian $W(\rho) = \sum_k \epsilon^k W_k$ while
$\Omega_j$ is the $j$-th order contribution to the following expression
$$
1-\frac{\omega_0}{\omega} =
\sum_{l=1}^{\infty} \frac {\epsilon^l \omega_l}{\omega_0} -
\sum_{l,l'} \frac {\epsilon^{l+l'} \omega_l \omega_{l'} }{\omega_0^2} +
\sum_{l,l',l"} \frac {\epsilon^{l+l'+l"} \omega_l \omega_{l'} \omega_{l"} }
{\omega_0^3} - \dots
$$
The static mean-field solution $\rho_0$ of Eq. (\ref{1.5}) corresponds to
$k=0$ in the notations of Eq. (\ref{2.4}).
It satisfies the static equation  $[W_0,\rho_0] = 0 $ where $W_0$ is the
static mean-field Hamiltonian. We define hole ($h$) and particle ($p$) states
(normalized to one) by
$$
\rho_0\vert \phi_h\rangle = \vert \phi_h\rangle, ~\rho_0\vert \phi_p\rangle =
0,
{}~W_0 \vert \phi_h\rangle = e_h\vert \phi_h\rangle,
{}~W_0 \vert \phi_p\rangle = e_p\vert \phi_p\rangle  ~ .$$

In first order one recovers the well known linearized
mean-field (RPA) equation
$ \hbox{i} \dot \rho_1 = {\cal M} \rho_1 $ which is solved by the ansatz
(\ref{2.3}). The RPA matrix ${\cal M}$ is defined by
$$
{\cal M} \rho_1 = [W_0,\rho_1] + [W_1,\rho_0] ~.
$$
Since $\rho(t)$ must correspond to a Slater determinant, it must satisfy
the condition $\rho^2(t) = \rho(t)$ to each order in $\epsilon$.
In $k$-th order $(k=0,1,2,..)$, this condition reads
\begin{equation} \label{2.5}
\rho_k = \sum_{j+m=k} \rho_j \rho_m ~.\end{equation}
An immediate consequence of this last equation is that it provides at once the
particle-particle and hole-hole matrix elements of the operator $\rho_k$
from the knowledge of the lower order terms $\rho_{k-1}, \rho_{k-2}, ...$
After checking that the matrix elements
$\langle p\vert \rho_k\vert p'\rangle $ and
$\langle h\vert \rho_k\vert h'\rangle $ given by Eq. (\ref{2.5}) are consistent
with the corresponding evolution equations (\ref{2.4}), one sees that the
information
contained in Eq. (\ref{2.4}) concerns only the particle-hole and hole-particle
matrix elements of $\rho_k$.

The corrections $\omega_l ~(l=1,2, ...)$ to the harmonic frequency $\omega_0$
are determined by requiring that there is no resonant term with frequency
$\omega_0$ in the evolution equation of the particle-hole matrix elements of
$\rho_{l+1}$. We wish to emphasize that this procedure is sometimes
inapplicable. Indeed, resonant terms with frequencies equal
to $2\omega_0,~3\omega_0, ...$ may appear in the construction of higher-order
terms when these frequencies already belong to the continuous spectrum of the
RPA matrix.
However, one can circumvent this difficulty by generalizing the ansatz
(\ref{2.3}). The treatment of resonant terms is
developped in Refs. \cite{16,21}.

Once the periodic orbits are obtained, they can be used to construct energy
spectra of $N$-body systems. The standard procedure is to perform a
semiclassical quantization of these orbits \cite{22}. This prescription
consists in selecting those solutions for which the mean-field action $I$
along a periodic orbit $\rho(t)$,
\begin{equation} \label{2.6}
I = \sum_h \int_0^T \hbox{d} t~
\langle \phi_h(t)\vert \hbox{i} \partial_t - \frac {\theta_h}{T}
\vert \phi_h(t) \rangle \end{equation}
is equal to an integer multiple of Plank's constant (or in
atomic units, $I = 2 \pi n$). The summation runs over the quasi-periodic
single-particle occupied states and $\theta_h$ is the so-called
Floquet-Lyapounov phase \cite{16}.

We now apply the method discussed above to the case of collective oscillations
of metallic clusters (see Eqs. (\ref{1.2}) and (\ref{1.6}) ).
In order to be able to discuss the two-phonon states,
we perform the calculations up to the third order in the amplitude of the
orbits. The explicit expressions of the contributions $W_k (k=0,1,2,3) $
to the mean-field Hamiltonian $W$ appearing in Eq. (\ref{2.4}) are
\begin{equation} \label{2.7} \begin{array} {ll}
W_0({\bf r}) & {\displaystyle =
-\frac{1}{2} \Delta + F_0({\bf r}) +
\int \hbox{d}^3 {\bf r'} \frac{1}{\vert {\bf r} - {\bf r'}\vert }
\rho_0({\bf r'}) + V_J (r)  }\\ \\
W_1({\bf r},t) & {\displaystyle =
\frac{ \hbox{d} F_0}{\hbox{d} \rho_0} \rho_1({\bf r},t) +
\int \hbox{d}^3 {\bf r'} \frac{1}{\vert {\bf r} - {\bf r'}\vert }
\rho_1({\bf r'},t) }\\ \\
W_2({\bf r},t) & {\displaystyle =
\frac{ \hbox{d} F_0}{\hbox{d} \rho_0} \rho_2({\bf r},t) +
\int \hbox{d}^3 {\bf r'} \frac{1}{\vert {\bf r} - {\bf r'}\vert }
\rho_2({\bf r'},t)  + \frac{1}{2}
\frac{ \hbox{d}^2 F_0}{\hbox{d} \rho_0^2} \rho_1^2({\bf r},t) }\\ \\
W_3({\bf r},t) & {\displaystyle =
\frac{ \hbox{d} F_0}{\hbox{d} \rho_0} \rho_3({\bf r},t) +
\int \hbox{d}^3 {\bf r'} \frac{1}{\vert {\bf r} - {\bf r'}\vert }
\rho_3({\bf r'},t)  +
\frac{ \hbox{d}^2 F_0}{\hbox{d} \rho_0^2} \rho_1({\bf r},t) \rho_2({\bf r},t)
+ \frac{1}{6}
\frac{ \hbox{d}^3 F_0}{\hbox{d} \rho_0^3} \rho_1^3({\bf r},t) }
\end{array} \end{equation}
The quantity $F_0 ({\bf r})$ in this last equations is
\begin{equation} \label{2.8}
F_0({\bf r}) = {\displaystyle
- (\frac{3}{\pi})^{1/3} \rho_0^{1/3} ({\bf r}) ~- 0.0333 ~\hbox {Ln}
[ 1 + 11.4 (\frac {4\pi}{3} \rho_0({\bf r}) )^{1/3} ]  } \end{equation}
and the notation $\rho_k({\bf r},t) ( k = 0,1,2,3 ) $ means
$\langle {\bf r}\vert \rho_k(t)\vert {\bf r}\rangle $.

The RPA matrix ${\cal M}^{(\lambda)}$ corresponding to the angular-momentum
$\lambda$ can be written as (see, e.g., Ref. \cite{20}) :
\begin{equation} \label{2.9}
{\cal M}^{(\lambda)} = \Bigg ( \begin{array} {ll}
{}~~~A^{(\lambda)} ~ ~~~~B^{(\lambda)}
\\ -B^{*(\lambda)} ~ -A^{*(\lambda)} \end{array} \Bigg )
\end{equation}
where the matrix $A^{(\lambda)}$ and $B^{(\lambda)}$ are
\begin{equation} \label{2.10} \begin{array} {ll}
A^{(\lambda)}_{\alpha_p\alpha_h\alpha{p'}\alpha{h'}} = &{\displaystyle
\delta_{\alpha_p\alpha{p'}} \delta_{\alpha_h\alpha{h'}} (e_p - e_h)
+ B^{(\lambda)}_{\alpha_p\alpha_h\alpha{p'}\alpha{h'}} } \\ \\
B^{(\lambda)}_{\alpha_p\alpha_h\alpha{p'}\alpha{h'}} = &{\displaystyle
\frac{1}{2\pi}
f^{(\lambda)}_{\alpha_p\alpha_h} f^{(\lambda)}_{\alpha_{p'}\alpha_{h'}}
\int_0^{\infty} \hbox{d}r ~r^2
R_{\alpha_p}(r) R_{\alpha_h}(r) ~\times } \\
&{\displaystyle ~~~~~~~~~~~~~~~~~~~~~~~~~~~~~~\big (
\frac{ \hbox{d} F_0}{\hbox{d} \rho_0} R_{\alpha_{p'}}(r) R_{\alpha_{h'}}(r) +
\frac{4\pi}{2\lambda +1} {\cal G}^{(\lambda)}_{\alpha_{p'}\alpha_{h'}} (r)
\big ) } \end{array}  ~.\end{equation}
In equations (\ref{2.10}) we have introduced the notation $\alpha_a \equiv
n_a,l_a $ where $n_a$ is the principal quantum number and $l_a$ the orbital
angular momentum of the single particle static wave function
$\langle {\bf r}\vert \phi_a\rangle $.
The quantities $e_a$ and $R_{\alpha_a}(r)$ are respectively the single particle
energies and the radial part of $\langle {\bf r}\vert \phi_a\rangle $. The
coefficient $f$ is given by
$$
f_{\alpha_p\alpha_h}^{(\lambda)} =
\sqrt {(2l_p+1)(2l_h+1)}
{}~\pmatrix{ l_p& l_h& \lambda \cr 0&0&0 \cr } ~~
$$
while $F_0(r)$ is already defined in Eq. (\ref{2.8}). The function ${\cal G}$
appearing in these equations is :
$$
{\cal G}^{(\lambda)}_{\alpha_p \alpha_h} (r) =
\int_0^{\infty} \hbox{d} r' ~r'^2 ~\frac {r^{\lambda}_<}{r^{\lambda+1}_>}
{}~R_{\alpha_p}(r') R_{\alpha_h}(r')
$$
where $r_<$ stands for Min$(r,r')$ while $r_> =$ Max$(r,r')$.

\renewcommand{\theequation}{\arabic{section}.\arabic{equation}}
\setcounter{equation}{0}
\section { Results }
 In this section we present some results obtained with the above formalism for
monopole, dipole, quadrupole and octupole modes in charged $Na^+_{21}$.
Calculations were performed on a lattice including 100 mesh points whose total
size is 30 a.u., i.e., $15.9 \AA$. In a first step we have solved the static
equations. The values we have found for the binding energy, radius and single
particle energies of
$Na^+_{21}$ are given in Table I. These values are in excellent agreement
with those obtained by Catara, Chomaz and Van Giai \cite{18} who use the same
energy functional but a different numerical method.
In a second step we have performed RPA calculations. The results for
the energies of collective
states of multipolarities $L=0,1,2,3$ and the corresponding percentages of the
sum rule and energy weighted sum rule are collected in Table II.
These values also agree well with those of reference \cite{18} for collective
modes. It can be noted from
Table II that dipole and quadrupole modes exhibit a strong collective
character while such is not the case for octupole and especially monopole
modes. For the
monopole mode one observes two neighbouring states with similar sum rule
fractions. The monopole mode is a nearly unbound RPA state while other modes
are bound. However, two-phonon states would belong to the continuum in the
harmonic approximation. Transition densities of dipole and quadrupole states
are displayed on Figs. 1 and 2 . The fact that these modes are bound is
reflected by the rapid decrease of these densities at large distance.
For reference the static electron density is shown on the same scale
(solid line).

The observation of the dipole resonance in alkali-metal clusters including
the charged $Na^+_{21}$ has been reported by several experimental groups, e.g.,
 Br\'echignac {\it et al} \cite{177} and J. Pedersen {\it et al} \cite{178}
(see also Ref. \cite {30} for a review).
A pronounced peak in the photoabsorption cross section is seen at
$\approx 2.7$ eV. This observation agrees with the RPA calculations
of Refs. \cite {9,11}. These authors predict the dipole mode in $Na^+_{21}$ at
$\approx 3.0$ eV with $\approx 80 \%$ of the sum rule and $\approx 83 \%$
of the energy weighted sum rule. These results also agree well with ours.

In a third step we have determined the location of the not yet observed
two-phonon states by
performing a semiclassical quantization of our second order periodic orbits as
prescribed by Eq. (\ref{2.6}). This prescription has been supplemented
by an approximate angular momentum projection which generates a splitting of
two-phonon states with different $J$'s \cite{16,21}. For most states this
procedure produces results which are stable with respect to  the
lattice size. The corresponding results are shown in Table III (a). One
observes
that small anharmonicities are found except for the $J=0$ two-octupole phonon
state for which a 15 percent deviation is obtained.

For the other states ( double monopole, $J=2$ double dipole and $J=0,2,4$
double quadrupole phonons) we were not able to calculate the energy shifts
by the method described in Sec. II because resonant terms appear during the
construction of second order terms.
For example, the $J=0$ two quadrupole phonon state couples with a resonant
mode of the RPA($L=0$) matrix so that the inhomogeneous linear equation
(\ref{2.4}) is singular in second order.
To study these states we use the modified perturbative construction
procedure presented in Ref. \cite{16}. As first order periodic orbit we take a
superposition of the $L$-mode with energy $\hbar \omega_L$ and the resonant
RPA mode with energy $\hbar \omega_{L_r} =  2\hbar \omega_L$
instead of the ansatz (\ref{2.3}).
With this modification, the semiclassical quantization procedure generates
not only the splitting of the two-phonon states ($n=2$) but also a splitting of
each state $n, J$ \cite{16}. The results are shown in Table III(b).
We have checked that the results in this table change
by less than 5 percent when the size of the lattice is increased to 128 mesh
points or decreased to 75 mesh points. One remarks first that the
energy shifts are small and compatible with zero so that
the anharmonicities of these states are negligible, second that for some states
the energy shift acquires an imaginary part.
We interpret this result as the signature of an instability of the
corresponding orbit.

A comparison with the results of Catara-Chomaz-Van Giai \cite{18} for the modes
in Table III is not straightforward. Indeed these authors find large
anharmonicities arising from couplings between states such as
$|(0^+,0^+)0^+>$ and $|(2^+,2^+)0^+>$ which are out of reach of our
semiclassical quantization approach. Further studies of this question are now
under way.

\renewcommand{\theequation}{\arabic{section}.\arabic{equation}}
\setcounter{equation}{0}
\section { Conclusion }

In this paper we have investigated the static and dynamic properties
of the alkali-metal cluster $Na_{21}^+$ in the jellium model.
Our results for the static and RPA calculations agree
with those of reference \cite{18}.
We have determined the splitting of the two-phonon states of multipolarities
$L=0, 1, 2, 3$  in charged $Na_{21}^+$.
This was achieved by using a perturbative method of constructing
periodic orbits of the mean-field equations recently developped
for giant resonances in nuclei \cite{16}.
In most cases, we find small anharmonicities.
To study the two-phonon states of the monopole and quadrupole modes we had to
use a more general method which
takes into account the coupling between the two-phonon state and the
resonant modes \cite{16}.
It turns out that these resonant terms generate anharmonicities which are
so small as to be compatible with zero.
For the $J=2$ two-dipole, $J=2$ two-quadrupole and $J=4$ two-quadrupole
states, we find an imaginary value for the energy shift with
respect to the harmonic two-phonon energy. These values are however small and
compatible with zero but are nevertheless a signature of an instability
of the corresponding orbit. Such instabilities cannot be detected by other
methods based on quasi boson expansions \cite{18}.

Finally, we wish to emphasize that periodic orbits provide a powerful
method to investigate collective energy spectra. This method has the
advantage of being rather easy to implement.
It was first found useful in the context of giant collective modes in nuclear
physics \cite{15,16} and also in the context of baryonic
physics to describe the Roper resonance \cite{17}. In both cases small
anharmonic terms were found. This seems to be true also for giant
collective modes in metallic agregates. This is a fortunate situation since
expansions in the elongation of periodic orbits often break down, even for
simple systems, because of the appearence of branches or bifurcations at small
values of the elongation \cite{23}. Here in contrast the region of interest
corresponding to one and two-phonon excitations lies well within the radius of
convergence of the series.

\vskip 1cm
{\large {\bf Acknowledgments} }

We wish to thank E. Lipparini for a fruitful discussion which motivated this
work.  We acknowledge discussions with F. Catara and N. Van Giai concerning the
comparison with our method.
We express our appreciation to H. Flocard and N. Pavloff for a critical
reading of the manuscript.
One of us, A. Abada, is grateful to N. Pavloff for his remarks, comments
and permanent interest in this work.
The Division de Physique Th\'eorique is " Unit\'e de Recherche des
Universit\'es Paris XI et Paris VI associ\'ee au CNRS".

\newpage
{\Large {\bf Table captions } }
\vskip 1cm

 {{\bf TABLE I.}}  ~Static properties of charged $Na^+_{21}$.
The binding energy $E_{LDA}$ is in eV, the root-mean-square radius $r_0$
in $\AA$ and the single particle energies are in eV.

\begin{tabular}{ c c c c c c c}
\\ \hline\hline
$E_{LDA}$ & $r_0$ & $1s$ & $1p$ & $1d$ & $2s$ & $1f$ \\ \hline
$-38.69$ & $4.49$ & $-7.55$ & $-6.82$ & $-5.83$ & $-5.15$ & $-4.64$\\
\hline\hline

\end{tabular}

{{\bf TABLE II.}} ~Low-lying RPA states of charged $Na^+_{21}$.
For each multipolarity $L^{\pi}$ we report the collective energy
$\hbar \omega_0$ (in eV) and the corresponding percentages of the sum rule
(SR) and energy weighted sum rule (EWSR).

\begin{tabular}{ c c c c }
\\ \hline\hline
$L^{\pi}$ & $\hbar \omega_0$ & $\%$ SR  & $\%$ EWSR  \\ \hline

$0^+$ & $4.5$ & $23$ & $20$ \\
      & $5.17$ & $22$ & $22$ \\
$1^-$ & $3.04$ & $87$ & $84$ \\
$2^+$ & $3.67$ & $55$ & $54$ \\
$3^-$ & $4.14$ & $33$ & $36$ \\ \hline\hline

\end{tabular}

\newpage

{{\bf TABLE III.}} ~Two-phonon states in charged $Na^+_{21}$.
For each state are reported its spin and parity and the shift $\delta E$
(in eV) with respect to the unperturbed two-phonon energy
$E^*_0 = 2\hbar \omega_0$ (in eV)
(calculated with semiclassical quantization [See Eq. (\ref{2.6})]).

\begin{center} {
\begin{tabular}{ c c c c }
{\bf (a)}
\\ \hline\hline
$L^{\pi}\otimes L^{\pi}$ & $ E_0^*$ & $J^{\pi}$ & $\delta E$ \\ \hline

$1^-\otimes 1^-$ & $6.08$ & $0^+$ & $-0.21$ \\
\hline
$3^-\otimes 3^-$ & $8.27$ & $0^+$ & $-1.24$ \\
                 &    & $2^+$ & $-0.52$ \\
                 &    & $4^+$ & $-0.20$ \\
                 &    & $6^+$ & $-0.15$ \\
\hline\hline
\end{tabular}
\hskip 2cm
\begin{tabular}{ c c c c }
{\bf (b)}
\\ \hline\hline
$L^{\pi}\otimes L^{\pi}$ & $ E_0^*$ & $J^{\pi}$ & $\delta E$ \\ \hline

$0^+\otimes 0^+$ & $8.98$ & $0^+$ & $\pm 0.01$ \\
$0^+\otimes 0^+$ & $10.34$ & $0^+$ & $0.00$ \\
\hline
$1^-\otimes 1^-$ & $6.08$ & $2^+$ & $\pm i 0.07$ \\
\hline
$2^+\otimes 2^+$ & $7.34$ & $0^+$ & $\pm 0.01$ \\
                 &    & $2^+$ & $\pm i 0.006$ \\
                 &    & $4^+$ & $\pm i 0.024$ \\
\hline\hline
\end{tabular}
} \end{center}

\vskip 1cm
{\Large {\bf Figure captions } }
\vskip 1cm

{\bf FIG. 1.} Static density $\rho_0(r)$ ($\AA ^{-3}$) (full line)
and transition density $\rho_1(r)$ ($\AA ^{-3}$) (dashed line)
corresponding to dipole state $1^-$ in the case of charged $Na^+_{21}$.

{\bf FIG. 2.} Static density $\rho_0(r)$ ($\AA ^{-3}$) (full line)
and transition density $\rho_1(r)$ ($\AA ^{-3}$) (dashed line)
corresponding to quadrupole state $2^+$ in the case of charged $Na^+_{21}$.

\end{document}